\documentclass[sigconf, nonacm]{acmart}
\usepackage{amsfonts}
\usepackage{booktabs}
\usepackage{pgfplotstable}
\usepackage{csvsimple}
\usepackage{siunitx}
\usepackage{xcolor}
\usepackage{xstring}
\usepackage{xspace}
\usepackage{enumerate}
\usepackage{enumitem}
\usepackage{algorithm}
\usepackage[noend]{algpseudocode}
\newcommand{\introparagraph}[1]{\medskip \noindent {\bf  #1.}}

\newcommand{\argmax}{\mathop{\mathrm{argmax}}}

\newcommand{\gpts}{Codex\xspace}
\newcommand{\Forum}{Forum\xspace}
\newcommand{\Product}{Product\xspace}
\newcommand{\JigsawM}{JigsawM\xspace}
\newcommand{\jc}[1]{\textcolor{pink}{[Joyce: #1]}}

\newcommand{\vl}[1]{\textcolor{brown}{[Vu: #1]}}
\newcommand{\ms}[1]{\textcolor{cyan}{[Mukul: #1]}}
\newcommand{\highlight}[1]{\textcolor{cyan}{#1}}

\newcommand{\eat}[1]{}
\newcommand\vldbdoi{XX.XX/XXX.XX}
\newcommand\vldbpages{XXX-XXX}
\newcommand\vldbvolume{14}
\newcommand\vldbissue{1}
\newcommand\vldbyear{2020}
\newcommand\vldbauthors{\authors}
\newcommand\vldbtitle{\shorttitle} 
\newcommand\vldbavailabilityurl{URL_TO_YOUR_ARTIFACTS}
\newcommand\vldbpagestyle{plain} 
\begin{document}
\title{From Words to Code: Harnessing Data for Program Synthesis from Natural Language}

\author{Anirudh Khatry}
\affiliation{%
  \institution{Microsoft}
}
\email{t-anikhatry@microsoft.com}

\author{Joyce Cahoon}
\affiliation{%
  \institution{Microsoft}
}
\email{joyce.cahoon@microsoft.com}

\author{Jordan Henkel}
\affiliation{%
  \institution{Microsoft}
}
\email{jordan.henkel@microsoft.com}

\author{Shaleen Deep}
\affiliation{%
  \institution{Microsoft}
}
\email{shaleen.deep@microsoft.com}

\author{Venkatesh Emani}
\affiliation{%
  \institution{Microsoft}
}
\email{venkatesh.emani@microsoft.com}

\author{Avrilia Floratou}
\affiliation{%
  \institution{Microsoft}
}
\email{avrilia.floratou@microsoft.com}

\author{Sumit Gulwani}
\affiliation{%
  \institution{Microsoft}
}
\email{sumitg@microsoft.com}

\author{Vu Le}
\affiliation{%
  \institution{Microsoft}
}
\email{levu@microsoft.com}

\author{Mohammad Raza}
\affiliation{%
  \institution{Microsoft}
}
\email{moraza@microsoft.com}

\author{Sherry Shi}
\affiliation{%
  \institution{Microsoft}
}
\email{shersh@microsoft.com}

\author{Mukul Singh}
\affiliation{%
  \institution{Microsoft}
}
\email{singhmukul@microsoft.com}

\author{Ashish Tiwari}
\affiliation{%
  \institution{Microsoft}
}
\email{ashish.tiwari@microsoft.com}

\eat{
\author{Zhe Zuo}
\affiliation{%
  \institution{East China Normal University}
  \city{Shanghai}
  \country{China}
}
\email{firstname.lastname@ecnu.edu.cn}

\author{Donald Fauntleroy Duck}
\affiliation{%
  \institution{Scientific Writing Academy}
  \city{Duckburg}
  \country{Calisota}
}
\affiliation{%
  \institution{Donald's Second Affiliation}
  \city{City}
  \country{country}
}
\email{donald@swa.edu}
\endcomment}

\newcommand{\jjh}[1]{\textcolor{blue}{(JJH: #1)}}
\newcommand{\sd}[1]{\textcolor{red}{(SD: #1)}}

\begin{abstract}

Creating programs to correctly manipulate data is a difficult task, as the underlying programming languages and APIs can be challenging to learn for many users who are not skilled programmers. Large language models (LLMs) demonstrate remarkable potential for generating code from natural language, but in the data manipulation domain, apart from the natural language (NL) description of the intended task, we also have the dataset on which the task is to be performed, or the {\em{data context}}. Existing approaches have utilized data context in a limited way by simply adding relevant information from the input data into the prompts sent to the LLM. 

In this work, we utilize the available input data to execute the candidate programs generated by the LLMs and gather their outputs. We introduce semantic reranking, a technique to rerank the programs generated by LLMs based on three signals coming the program outputs: (a) \emph{semantic filtering and well-formness based score tuning}: do programs even generate well-formed outputs, (b) \emph{semantic interleaving}: how do the outputs from different candidates compare to each other, and (c) \emph{output-based score tuning}: how do the outputs compare to outputs predicted for the same task. We provide theoretical justification for semantic interleaving. We also introduce temperature mixing, where we combine samples generated by LLMs using both high and low temperatures. We extensively evaluate our approach in three domains, namely databases (SQL), data science (Pandas) and business intelligence (Excel’s Power Query M) on a variety of new and existing benchmarks. We observe substantial gains across domains, with improvements of up to $45\%$ in top-1 accuracy and $34\%$ in top-3 accuracy.
\end{abstract}

\maketitle

\eat{
\pagestyle{\vldbpagestyle}
\begingroup\small\noindent\raggedright\textbf{PVLDB Reference Format:}\\
\vldbauthors. \vldbtitle. PVLDB, \vldbvolume(\vldbissue): \vldbpages, \vldbyear.\\
\href{https://doi.org/\vldbdoi}{doi:\vldbdoi}
\endgroup
\begingroup

\renewcommand\thefootnote{}\footnote{\noindent
This work is licensed under the Creative Commons BY-NC-ND 4.0 International License. Visit \url{https://creativecommons.org/licenses/by-nc-nd/4.0/} to view a copy of this license. For any use beyond those covered by this license, obtain permission by emailing \href{mailto:info@vldb.org}{info@vldb.org}. Copyright is held by the owner/author(s). Publication rights licensed to the VLDB Endowment. \\
\raggedright Proceedings of the VLDB Endowment, Vol. \vldbvolume, No. \vldbissue\ %
ISSN 2150-8097. \\
\href{https://doi.org/\vldbdoi}{doi:\vldbdoi} \\
}\addtocounter{footnote}{-1}\endgroup

\endeat}

\section{Introduction}

The emergence of Large Language Models (LLMs), such as Codex~\cite{chen2021evaluating} and GPT~\cite{brown2020language, openai2023gpt4}, has fundamentally transformed the field of program synthesis from natural language (NL), leading to the rapid development of NL interfaces (i.e., ~\cite{PowerAutomate, qaPowerBI}) and code assistants~\cite{githubcopilot} that are widely used by practitioners and developers. In-context learning~\cite{dong2023survey} plays a crucial role in this transformation, enabling LLMs to generate code for a diverse range of programming languages with minimal input. The input to LLM typically consists of a prompt describing the task to be performed in natural language and potentially a few examples, which are used for few-shot prompting~\cite{brown2020language, logan2021cutting}. By leveraging in-context learning, LLMs can acquire domain-specific knowledge that enhances their understanding of the syntax and structure of the programming language that the generated code must adhere to. 
An LLM-based approach for NL to code tasks is especially valuable as it does not require training and deploying custom models for specific programming domains (e.g., SQL, Python) resulting in a more streamlined design.


The simplicity and strength of in-context learning makes it an excellent choice for synthesizing data manipulation programs, i.e. programs that are designed to extract, query, and transform data using NL descriptions. A crucial aspect that sets this problem apart is the need to take into account the \textit{data context}: in addition to the NL description of the intended task, we also have context about the dataset on which the task needs to be performed, such as the data schema and data values. While prior work has also utilized the data context, it is largely limited to using the data context for creating better prompts~\cite{synchromesh, pourreza2023dinsql, trummer2022codexdb}.
These prompts typically contain a description of the task along with information about the input data, such as the schema (column names) and sample rows. 

In this paper, we take a step further and show that the data context can be leveraged in more profound ways to enhance the performance of code generation tasks. In particular, we make the observation that given a program associated with a task and a sample of the input dataset, we can generate the output of the task by executing the program on this sample input. 
These task outputs (or the inability to construct them due to, for instance, generated code being malformed) provide valuable information that can be used in conjunction with task inputs during the code generation process, creating a rich space of new design possibilities.



\begin{figure*}[tb]
    \centering
    \includegraphics[width=0.9\linewidth]{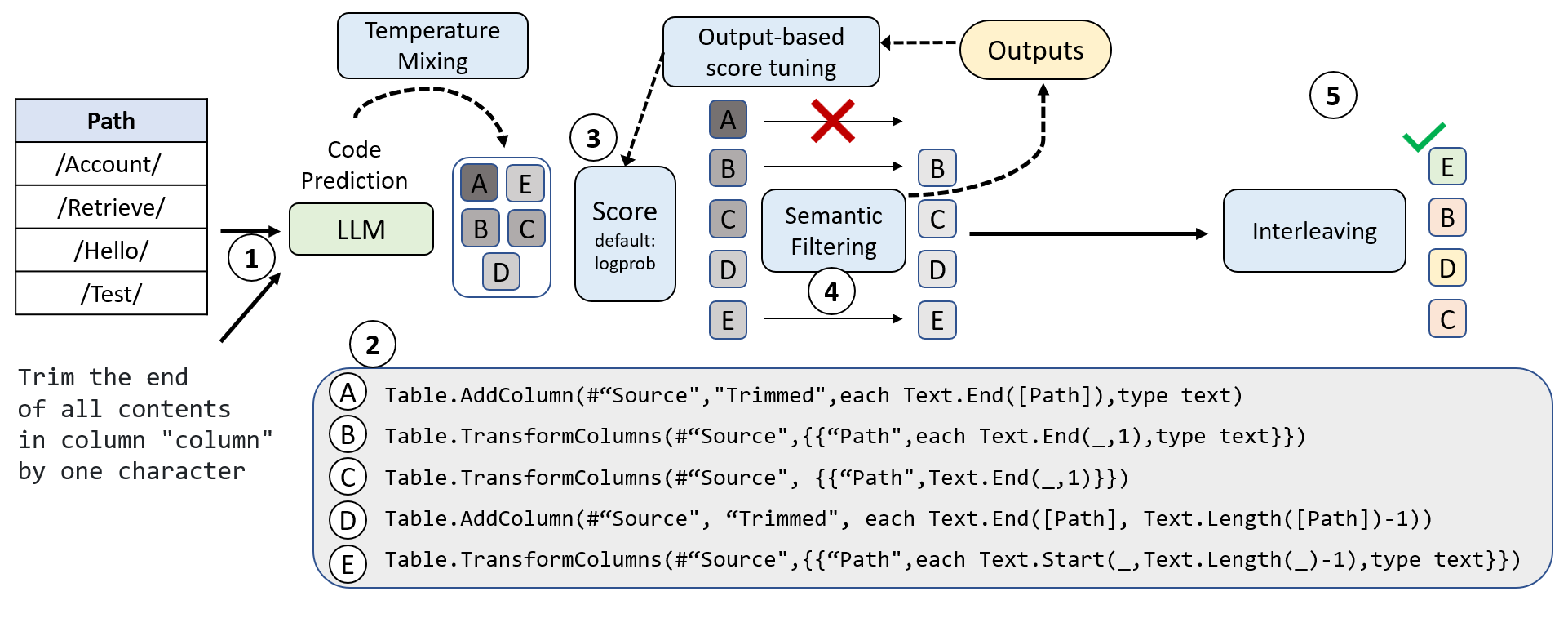}
    \caption{Architecture diagram highlighting the working of the system. (1) Data and natural language description are both provided to an LLM. (2) The LLM generates 25 samples, using a mix of samples generated using different temperatures, of which five are shown represented by A, B, C, D and E. (3) The candidates are ranked by score, which defaults to \textit{average logprobs}. (4) The candidates are executed, and non-executable programs are filtered. (5) The outputs are used to optionally tune the scoring function, candidates are clustered based their outputs, and the clusters are interleaved to generate a final ranked list. The dotted arrows indicate optional components.
    }
    \label{fig:architecture}
\end{figure*}

Our proposed architecture to leverage data context in advanced ways is shown in Figure~\ref{fig:architecture}. Instead of only feeding the LLM a data-enhanced prompt and accepting the LLM's top result, we request the top-$N$ results from the LLM to form a set of candidate programs, which are ranked by the probabilities of their corresponding syntactic tokens (logprobs\footnote{Informally, a higher logprob for an output generated by the LLM indicates that the model \emph{endorses} the output.}). We observe that although the correct program might be present within the set of candidate programs, it may not be at or near the top.
To surface the correct programs to the top, we re-rank the candidate programs using a novel technique called \textit{semantic reranking} consisting of two steps namely \textit{semantic filtering} and \textit{semantic interleaving}. Semantic ranking relies on candidate program outputs obtained by directly executing the programs on a sample of the input data. If code execution fails and throws an exception, then that candidate is removed from the set (semantic filtering). After applying semantic filtering, each remaining candidate has a corresponding output. We utilize these outputs to group the programs into classes and then re-rank them in a way that increases the likelihood of the correct program being ranked higher, thus boosting the top-$K$\footnote{$K$ is the number of answers that are surfaced to the user.} accuracy (semantic interleaving). \eat{\vl{I feel that the relation ship of these new terms are not very clear (esp. filtering and interleaving vs semantic ranking).}. {\color{red} AF: Rephrased it a bit. Vu please take a look}.} 

Semantic reranking improves accuracy of top-$K$ code predictions, but we observed that the gains were smaller for less common target languages. This is probably due to limited training data on these languages being available on the web. 
In such situations, we can use the capability of LLMs to directly generate the output dataset for a given task (\textit{output prediction}) from the NL task description and input dataset.
We then compare the predicted outputs with the output generated by execution of a predicted program, and use the result to refine the logprob score assigned to that program. This {\em{output-based score tuning}} provides the semantic reranker a better score function that leads to additional top-$K$ improvements.
We also introduce {\em{temperature mixing}} to overcome tradeoffs from using either very low or very high temperature when using an LLM.



\smallskip
\introparagraph{Our Contribution}  In this paper, we present a novel approach for synthesizing data manipulation programs that leverages the data context in previously unexplored ways to generate more accurate programs. We make the following contributions.
\begin{itemize}[noitemsep,topsep=0pt]
    \item We present a novel program synthesis framework for the data manipulation domain. Our framework leverages data context end-to-end by exploiting \emph{both} task inputs and outputs. At the core of our approach lies a new semantic ranking technique that takes into account the semantic diversity of the programs based on their outputs.
    \item We propose a novel technique to tackle scenarios where the LLM lacks sufficient prior knowledge of a specific data manipulation language. This technique leverages data context in a related LLM task (i.e., output prediction) and utilizes the results as inputs to the code generation task resulting in further accuracy improvements.
    \item We demonstrate the applicability of our approach in three domains, namely databases (SQL), data science (Pandas), and business intelligence (Excel's Power Query M~\cite{m}) using a variety of new and existing benchmarks. While SQL and Pandas are extremely popular languages for data retrieval and manipulation that are used in/by multiple vendors, M is used by several products in Microsoft that power business analytics. The LLM has likely seen much fewer examples of M programs compared to SQL and Pandas. This allows us to conduct analysis for data manipulation tasks for the scenario where the model has inadequate prior knowledge. Extensive empirical evaluation shows that for all three domains, we achieve substantial gains with improvements of up to $45\%$ on top-$1$ accuracy and $34\%$ on top-$3$ accuracy.
\end{itemize}

\section{Motivating Scenario}
\label{sec:motivation}

In this section, we motivate our solution using an example NL statement and a real-world use-case associated with Power Query M~\cite{powerqueryM} expression\footnote{Most of the observations and insights discussed in this section apply to SQL and Python as will be demonstrated by our experimental evaluation in Section~\ref{sec:eval}.}.
M is a data manipulation language used in business intelligence applications such as Microsoft's Excel, and PowerBI~\cite{powerBI} and is typically used to filter, transform and combine data from one or more sources. Several prior works~\cite{raza2020web, altwaijry2015query, zhu2017auto} within the database community have been motivated by M due to its capability to dynamically find, visualize, share, and query data across a wide variety of online and offline sources. \eat{We have chosen to present an example using M because it enables us to demonstrate some of the techniques we have developed for languages where the LLMs have insufficient prior knowledge due to insufficient training data.}

In our scenario shown in Figure~\ref{fig:architecture}, a user is working in Power Query with a table \texttt{Source} containing a column named
\texttt{`Path'} with some text data. 
The user issues a NL query: \textit{``Trim the end of all contents in column "Path" by one character''} and wants an M expression for it (this is a real query which was scraped from a Help Forum\footnote{https://stackoverflow.com/questions/72548765/trimming-end-of-column-in-power-query}). If the expression is correct, the user can proceed executing it on their data and collect the output.

We first prepare the best possible prompt for an LLM that includes this NL query, the schema of the table Source, some sample rows from Source, and some examples of NL queries alongside the corresponding M expressions (few-shot prompting).
We send the prompt to LLM and ask it to produce multiple candidate programs ($N=25$) with the associated M expression corresponding to this NL utterance.
Some of the candidates are shown in Figure~\ref{fig:architecture}. 
As shown in the figure, the top program $A$ produced by the LLM is syntactically well formed, but it is semantically incorrect as $\mathtt{Text.End}$ requires two arguments and not one (the second argument denotes the number of characters to be selected).

For this example, we observed that the correct program is in position $\#10$ (not shown). This is undesirable as the users can only be reasonably expected to see the first few candidates at most. An obvious solution would be to order the candidate programs by the average logprobs provided by the model. The LLM can return a probability associated to each token that composes the created text. These probabilities, returned in logarithmic form, measure how likely each token is to occur in the context of the output text. After ordering the candidate programs, the correct program is now ranked at position $\#5$ (shown as $E$ in Figure~\ref{fig:architecture}). This is an improvement but still unlikely to be sufficient for most applications. 

As we observed with Program $A$ in Figure~\ref{fig:architecture}, LLMs can return programs that are syntactically or semantically incorrect. 
These programs will produce errors when executed. A reasonable approach would be to remove all candidates that produce errors during program execution. After performing this \textit{semantic filtering} step (cf. Section~\ref{sec:filtering}), the correct solution moves to position $\#4$. In practice, this is a substantial improvement when there are multiple candidates but it is worth examining whether we can further advance the correct program. 

Since we executed the programs and have their outputs, a natural next step is to examine whether these outputs can be leveraged to improve the ranking of the candidate programs. Taking a closer look, we observe that the candidates $B$ and $C$ both return the column with only the last character.
Thus, the two programs may be semantically equivalent, and hence we could demote candidate $C$ to go below candidates $D$ and $E$ since they produce outputs that are {\em{different}} from those produced by candidates before them ($B$ and $C$). After this \textit{semantic interleaving} step (see Section~\ref{sec:interleaving}), the correct program is placed in position $\#3$.

A natural question that arises is whether there is any additional information that we can leverage to further improve the position of the correct program. Assuming there was an oracle that possessed advanced knowledge of the expected output data for the given NL statement, we could easily elevate the position of the program that generates that output to the top of the list. In the absence of an oracle, we can leverage the LLM itself for this task. In particular, we can ask the LLM to predict directly the data output (not the M code) that corresponds to the NL statement and the input table. Now, we can re-rank the candidate programs by leveraging this additional information. In particular, we notice that the output of candidate $E$ exactly matches the top output predicted by the LLM and this is not the case for candidates $B$ and $D$. We can thus promote candidate $E$ to the top of the list by refining the average logprobs-based score assigned to each candidate. After using this {\em{output-based score tuning}} (cf. Section~\ref{sec:output_prediction}), the correct answer is at the top.

Note that in the interest of communicating the major insight to the reader, we omitted discussing a few steps shown in Figure~\ref{fig:architecture} such as temperature mixing and score computation. These techniques will be discussed in detail in the following sections.

\section{Exploiting Data Context}
\label{sec:outline}



We begin by first describing the problem setup, followed by a high-level description of our overall approach.

\subsection{Problem Setup} \label{subsec:setup}

The problem we consider in this work can be stated as follows: Given a natural language description $nl$ of some task along with the dataset $D$ over which that task should be performed, our objective is to generate the expression or program $s$ in some desired target language that will accomplish the task.\eat{Rather than designing specialized solutions and models, we want to exploit general-purpose large language models (LLMs) for this task (for all target languages). } As we mentioned before, the so-called data context, which consists of the dataset $D$, is clearly an important source of information for generating the desired program $s$. The most common way of exploiting the data context is to include a summary of $D$ in the prompt that is passed to the LLM. This step is now standard \cite{brown2020language, chen2022program, chen2021evaluating, nijkamp2022codegen} and we do \emph{not} consider the problem of prompt generation in this paper. Instead, in all our experiments, we used the best possible prompt we could design for the task by leveraging the prompts proposed in the existing literature.

 As shown in Section~\ref{sec:motivation}, our focus is on scenarios where user participation is involved. Concretely, the user submits the natural language query $nl$, which then initiates the process of generating $K$ programs ($K \geq 1$) as potential answers to this query. Once the programs are generated, the user will review them to determine if any of them aligns with their intended outcome. If any of them does, the user may proceed to execute it on their dataset $D$ to obtain the desired output. Our goal is to improve the top-$K$ accuracy for the task described above, that is to ensure that the correct program is present in the $K$ programs shown to the user for a given NL query. We make the following observations regarding reasonable values of $K$ for our setup:
 \begin{itemize}
     \item \textbf{K $=$ 1}. While improving top-1 accuracy akin to prior work~\cite{gu2023, fu2023catsql} is important, the NL question posed by the user can be inherently ambiguous and thus surfacing only one result might not cover the user's intent. Additionally, users might have preferences regarding the type of programs they prefer to execute. In the context of SQL for example, a user might prefer executing a nested SQL query than a query that creates a view even if both produce the correct output. Therefore, it is desirable to consider $K > 1$.
     \item \textbf{K $\gg$ 1}. Prior work has evaluated code generation tasks for large values of $K$ (i.e., K = 100 in~\cite{chen2021evaluating}). However, in settings where a human observes the generated programs, returning such a large number of programs is not feasible.
     \item \textbf{K $\leq$ 3}. As shown by relevant product efforts targeting similar scenarios~\cite{PowerAutomate}, demonstrating 3 answers to the user is indeed feasible as it improves the chances the correct program will be surfaced to the user while helping address the inherent ambiguity in natural language. 
 \end{itemize}

Given the above considerations, we aim to improve the top-$K$ accuracy for small values of $K$. Our results are shown in Section~\ref{sec:eval}.

\introparagraph{LLM parameters} There are several tuneable parameters exposed by the LLM. The first parameter is the prompt which contains information about the task and the data context. The second parameter is the \emph{temperature} which is a value between $0$ and $1$ (both inclusive). A lower temperature value makes the model behavior more deterministic (i.e. less variability in the answers returned by the model for the same input), whereas a high temperature makes the model take more risks and be "creative" in its response. In other words, the temperature parameter controls the degree of ``randomness'' that would be added when the model is sampling from the output probability distribution. This behavior allows us to get a diverse set of responses from the LLM. The model API also provides a parameter \emph{N} that controls the number of responses to generate for each prompt in a single call to the model. We use this parameter to obtain $N=25$ candidate programs in our experiments. Finally, for each response generated, the LLM can also return a probability value associated with each token ($w_i$) of the response. These probabilities, returned in logarithmic form (logprobs~\cite{shi2022natural}), measure how likely each token is to occur in the context of the previous tokens and the given prompt. Lower log probabilities indicate less likely words, while higher log probabilities indicate more likely words. \eat{We refer the reader to~\cite{api} for a description of the model API.}

\subsection{Our Approach}

Our techniques apply in the post-processing phase -- that is, after the LLM has generated $N$ candidates -- as illustrated in Figure~\ref{fig:architecture}. Our overall approach is as follows, and is also outlined in pseudocode in Algorithm~\ref{algo:overall}: 
\begin{description}[leftmargin=3ex]
\item[Prompt generation]
The description $nl$ and a summary of the dataset $D$, typically consisting of the schema and a small sample of $D$, is put together into a prompt that also includes some few-shot examples of the task. 

\item[Temperature Mixing]
The LLM is asked to generate $N$ candidates using a high temperature setting to increase the diversity, but we also optionally add the top candidate generated with temperature 0 into the mix. Temperature mixing is discussed further in Section~\ref{sec:boosting}.

\item[Score calculation] Some LLMs provide logprobs for all the tokens in a candidate. We assign a score to each candidate based on the average of the logprobs of its tokens. 
We will later describe how we refine this score computation. We note that ranking the candidates based on their average logprob score is part of standard practice \cite{shi2022natural}. The list $L$ of $N$ candidates and the score function $\mathtt{score}$ is the starting point for the application of our proposed new techniques.

\item[Semantic reranking]
We rerank the $N$ candidates in $L$ using a technique termed {\em{semantic reranking}} discussed in Section~\ref{sec:reranking}. It consists of two steps:
\begin{description}[leftmargin=4ex]
 \item[Semantic filtering] Each of the $N$ candidates in $L$ is executed on the input dataset $D$ (or a small sample if $D$ is too large). If the execution fails and throws an exception, we remove that candidate from the ranked list $L$. The result is a new ranked list $L'$ of $N'$ candidates where $N' \leq N$. 
 Semantic filtering is further discussed in Section~\ref{sec:filtering}.

 \item[Semantic interleaving]
 The candidates in $L'$ are re-ordered to give a new ordered list $L''$ of $N'$ elements. The reordering is guided by the outputs generated by the $N'$ candidates. Candidates that generate the same output as some higher-ranked candidate are moved lower in the ordering. Essentially, we group the $N'$ candidates by their outputs, order each group by score, and then zip these groups to get the final result $L''$.  The idea is to increase diversity in the outputs generated by the top $K$ candidates in $L''$. Semantic interleaving is discussed is Section~\ref{sec:interleaving}. 
\end{description}

\item[Output consistency]
In this optional step described in Section~\ref{sec:output_prediction}, the score (that is used above during semantic reranking) assigned to a candidate $s$ is refined based on (a) the well-formedness of the output $o_{exe}$ produced by $s$ on input $D$, and (b) the similarity of $o_{exe}$ to the output $o_{pred}$ 
that the LLM predicts would be generated if the $nl$ task were to be performed on $D$ (or a sample of $D$). 
Intuitively, if output $o_{exe}$ is identical or close to $o_{pred}$, then the candidate's score improves, and if the output $o_{exe}$ is not well-formed, then the candidate's score is punished. 
Here the definition of ``well-formedness'' can be customized. For example, if $o_{exe}$ contains a new column of null values, or a new column with majority missing values, then $o_{exe}$ could be classified as being ill-formed. 

\end{description}



\begin{algorithm}[t]
  \small
  \begin{algorithmic}[1]
    \Require An NL description $nl$ of some task and an input dataset $D$
    \Ensure Return top $K$ candidate programs to perform task $nl$ on $D$
    \Function{NL2Code}{$nl$, $D$}
      \State $\mathsf{prompt} \gets \Call{PreparePrompt}{nl, D}$
      \State $L, \mathtt{logprobs} \gets \Call{LLM}{\mathsf{prompt}, \mathtt{tmp}=0.6, N=24}\; \cup$ \label{line:temp-mixing}
      \Statex $\qquad\qquad\qquad\qquad\Call{LLM}{\mathsf{prompt}, \mathtt{tmp}=0, N=1}$
      \State $L', O_{\mathtt{exe}} \gets \Call{SemFilter}{L, D}$
      \State $O_{\mathtt{pred}}, \mathtt{Ologprobs} \gets \Call{OutputPrediction}{nl, D}$
      \State $\mathtt{scores} \gets \Call{GetScores}{L',\mathtt{logprobs},O_{\mathtt{pred}}, \mathtt{Ologprobs}, O_{\mathtt{exe}}}$
      \State $L'' \gets \Call{SemInterleave}{L', O_{\mathtt{exe}}, \mathtt{scores}}$
      \State \Return First $K$ candidates from $L''$
    \EndFunction
  \end{algorithmic}
  \caption{Our overall approach for generating top $K$ candidate programs given NL description $nl$ and dataset $D$.}
  \label{algo:overall}
\end{algorithm}

\section{Semantic Reranking}

\label{sec:reranking}

As described in the previous section, we prompt the LLM to obtain $N$ candidate programs. We often observe that the correct program is indeed included in the set of candidate programs returned, but it is ranked lower, making it challenging to surface. The scenario discussed in Section~\ref{sec:motivation} shows an example of such a case. 
The goal of semantic re-ranking is to bring the candidates that are most likely to be the desired programs to be near the top of the list so that if we surface only a small number of $K$ candidates to the user, we have a high likelihood of the desired program being included.



The reader may wonder why the default order produced by LLMs performs poorly in many cases. 
The main observation here is that the $N$ random variables that correspond to the $N$ program samples returned by the LLM are {\em{independent and identically distributed (i.i.d) random variables}}. Each candidate is sampled from the same underlying probability distribution, as computed by the model based on the NL input, hence identically distributed. (Here the set of possible programs is the probability space, and LLMs are viewed as generating a probability distribution on this space conditioned on a specific NL utterance.) Moreover, sampling is performed with replacement, since the same candidate can appear multiple times in the $N$ samples. Hence, each program sample is independent. Thus, LLMs generate i.i.d samples. 

However, to surface the most relevant top-$K$ candidates to the user, we want the $K$ samples to depend on each other. In fact, they should cover (as much as possible) the space of all possible programs implied by the (ambiguous) NL description.
Specifically, we want the $K$-th sample to be the most probable sample {\em{given that the other samples are not the intended ones}}. Thus, the choice of the 2nd candidate, for example, should be guided by the fact that the 1st candidate was not intended. The goal of semantic re-ranking is thus to increase the probability of surfacing the desired program in the top-$K$ candidates.\eat{

}
The Semantic Ranking algorithm relies on the available data context to re-rank the candidate programs. As shown in Algorithm~\ref{algo:semantic-reranker}, it consists of two steps: semantic filtering and semantic interleaving, which we further elaborate on below. 
\begin{algorithm}[t]
  \algloop{try}
  \algloop{catch}
  \small
  \begin{algorithmic}[1]
    \Require An array $L$ of $N$ candidate programs, an input dataset $D$, and a function $\mathtt{score}$ that assigns a score to each candidate in $L$
    \Ensure A reordering $\mathsf{rankedL}$ of a subset of $L$
    \Function{SemFilter}{$L$, $D$}
      \State $d \gets$ Sample of $D$
      \For{$i \in \{ 0, \ldots, N-1\}$}
        \try
          \State $O[i] = $ exec $L[i]($d$)$ \Comment{Run $L[i]$ on $d$ and save output}
        \catch{} \Comment{Execution threw an exception}
            \State Remove $L[i]$ from array $L$
      \EndFor
      \State \Return $L$, $O$
    \EndFunction
    \Function{SemInterleave}{$L$, $O$, $\mathtt{scores}$}
      \State $L \gets \mathtt{Sort}(L,\mathtt{scores})$
      \State $\mathsf{OrderedClasses} \gets \mathtt{new}\,\mathtt{Array}[][]$
      \State $\mathsf{ClassRepresentative} \gets \mathtt{new}\,\mathtt{Array}[]$
      \For{$i \in \{ 0, \ldots, \mathtt{len}(L)-1\}$}\label{line:eq_class_start}
        \If{$O[i]$ is $\mathsf{ClassRepresentative}[j]$ for some $j$}
           \State Add $L[i]$ to the end of array $\mathsf{OrderedClasses}[j]$
        \Else
           \State $j = \mathtt{len}(\mathsf{ClassRepresentative})$ \Comment{Create new class}
           \State $\mathsf{ClassRepresentative}[j] \gets O[i]$
           \State $\mathsf{OrderedClasses}[j] \gets [L[i]]$ \Comment{A new FIFO queue}\label{line:eq_class_end}
        \EndIf
      \EndFor
      \State {$\mathsf{RankedL} \gets []$}
      \While {some program remains in $\mathtt{OrderedClasses}$} \label{line:interleave_start}
        \For{$i \in \{ 0, \ldots, \mathtt{len}(\mathsf{ClassRepresentative})-1\}$}
          \If{$\mathtt{OrderedClasses}[i]$ is nonempty}
            \State {Pop $\mathtt{OrderedClasses}[i][0]$, add to end of $\mathsf{RankedL}$}\label{line:interleave_end}
          \EndIf
         \EndFor
      \EndWhile
      \State \Return $\mathsf{RankedL}$
    \EndFunction
  \end{algorithmic}
  \caption{The semantic reranker}
  \label{algo:semantic-reranker}
\end{algorithm}

\subsection{Semantic Filtering}
\label{sec:filtering}

The availability of the data $D$ provides the option to execute the code snippets generated by the LLM on that data. If executing a program produces errors, that program can be removed.
This is our first step in re-ranking the candidates generated by the LLM. 

Let $s$ be a candidate program sample generated by the model. We execute $s$ on a small sample of the input dataset $d \subseteq D$. The sample $d$ is drawn in a way that takes into account potential PK/FK relationships so that when we execute join operations, we can have meaningful results. If the execution is successful (i.e., completes without throwing an exception), then we say that the program $s$ is \emph{semantically valid}. In the semantic filtering step, we only keep the candidate programs that are semantically valid. 
The pseudocode for semantic filtering is shown in Algorithm~\ref{algo:semantic-reranker} as function {\textsf{SemFilter}}. Note that we execute all the candidate programs on the same input data sample.

Semantic filtering helps in two cases: (1) identify candidate programs that are syntactically wrong (i.e., do not conform to the syntax of the programming language), and (2) identify programs that are syntactically correct but reference tables and columns that do not exist in the data or use non-existent operators, which may be generated due to model hallucinations. 
%

An example is shown in Figure~\ref{fig:architecture} where the M program $A$ is semantically filtered out as it results in errors when executed. As we demonstrate in Section~\ref{sec:eval}, by leveraging this technique as a baseline filter, we can significantly improve the quality of the top-$K$ results from the LLM.

\subsection{Semantic Interleaving}
\label{sec:interleaving}

After applying semantic filtering, we obtain a subset of programs that we know are well-structured and meaningful. The main challenge then becomes surfacing a \textit{diverse} set of programs from this subset to the user as the programs that are left oftentimes have minor syntactic variations, but are semantically equivalent. We do this by classifying each program into equivalence classes, and selecting programs appropriately from different equivalence classes.

First, we use a commonly used approach that performs a ranking of the candidates based on the logprobs associated with each token ($w_i$) of the program sample. \eat{The LLM can return a probability value associated with each token ($w_i$) of a program sample. These probabilities, returned in logarithmic form (logprobs), measure how likely each token is to occur in the context of the previous tokens and the given prompt. Lower log probabilities indicate less likely words and higher log probabilities indicate more likely words.}
In particular, we use these probabilities to rank the candidate programs by assigning a ``score'' ($\bar{p}$) for each candidate, which is computed by taking the average of the conditional log probabilities of the tokens in each program candidate that is generated by the LLM. In other words:  
$$
\bar{p}_s = \frac{1}{n_s} \sum_{i=1}^{n_s} \log Pr(w_{s,i} | nl, w_{s,1}, \ldots, w_{s, i-1}),
$$
where $n_s$ is the number of tokens for a generated program sample $s$ in $N$ and $w_{s,i}$ represents the $i$-th token in program candidate $s$. 
We leverage previous works that propose using the average log likelihood as a selection criterion among samples \cite{chen2021evaluating, shi2022natural}. 
This average logprobs ``score'' is used as a ranking measure, with higher score driving the program higher up the rank. 

There are two issues with using this heuristic for ranking and selecting top $K$ candidates. 
One is that it can be unfairly impacted by the length of the candidate. For example, if a candidate is long and contains several {\em{predictable}} tokens, they can skew the average logprobs toward higher values. For example, consider the token {\texttt{avg}} that occurs in some program candidate. The token predicted to follow this token will be {\texttt{"("}} with very high probability. This predictable token  {\texttt{"("}} can substantially increase the average logprob of this candidate.
If a candidate contains many such predictable tokens, then its average logprob score will be higher, whether or not this candidate is actually reflective of the desired program. 
Even if we assume that all candidates have about the same number of predictable tokens and average logprobs provide a good surrogate for a candidate's probability of being the intended program, there are still issues with this approach which we extensively discuss in Section~\ref{sec:justification}.


The pseudocode for semantic interleaving is given in Algorithm~\ref{algo:semantic-reranker}. Given a list $L$ of candidates, their outputs $O$ as produced by the semantic filtering step, and a $\mathtt{score}$ function, we first rank the candidates in $L$ using the $\mathtt{score}$ function. Let $L$ be the ranked list; that is,
$\mathtt{score}(L[i]) > \mathtt{score}(L[j])$ for every $j > i$. Let $O[i]$ denote the result of executing $L[i]$ on the dataset sample $d$. We first partition candidates in $L$ into equivalence classes based on their outputs. 
Let $s_j$ denote the $j$-th candidate $L[j]$ in the list $L$. 
The equivalence class $[s_i]$ of $s_i$ is defined as:
\begin{eqnarray*}
 [s_i] & = & \{s_j \in L \mid O[j] = O[i] \}
\end{eqnarray*}
The equivalence classes are built on Lines~\ref{line:eq_class_start}-~\ref{line:eq_class_end} in Algorithm~\ref{algo:semantic-reranker} where they inherit ordering from $L$; that is, the score of an equivalence class is the score of its highest-ranked member.
\begin{eqnarray*}
 [s_j] > [s_i] & \mbox{if} & \mbox{there exists $s\in [s_j]$ s.t. $\mathtt{score}(s) > \mathtt{score}(s')$} 
 \\ & & \mbox{for all $s'\in [s_j]$}
\end{eqnarray*}
Let $[s_1] > [s_2] \cdots > [s_m]$ be the $m$ equivalence classes we obtain from candidate list $L$ ordered by the above ordering.
Let the class $[s_i]$ contain the candidates $s_{i,1}, s_{i,2}, s_{i,3}, \ldots$ ordered by $\mathtt{score}$. Then, the result of semantic interleaving  is the following ordered set of candidates:
$$
 s_{1,1}, s_{2,1}, \ldots, s_{m,1}, s_{1,2}, s_{2,2}, \ldots, s_{m,2}, s_{1,3}, s_{2,3}, \ldots, s_{m, 3}
$$
In other words, we consider the equivalence classes in order, and we pick the top members from each equivalence class, then the second best member from each equivalence class, and so on. 
This interleaving of equivalence classes is carried out on Lines~\ref{line:interleave_start}-~\ref{line:interleave_end} in Algorithm~\ref{algo:semantic-reranker}.

\subsubsection{Justifying Semantic Reranking}
\label{sec:justification}

Let $PS$ be the space of all programs given user natural language $nl$ description and the given data context $D$. 
For any program $s\in PS$, let $Pr(s \mid nl, D)$ denote the probability of the user accepting $s$ 
given the NL description $nl$ and input dataset $D$.
When we use a large language model to generate code, the model (internally)
generates a probability distribution $Pr$ on program space $PS$ conditioned on its prompt, and let us assume
that the LLM generates $Pr(\_\mid nl, D)$.

For simplicity, let us say we want to present only $K=2$ suggestions to the user.
The argument will generalize to any $K$. To maximize the chance that the user finds their intended program in the $K=2$ programs
presented to them, we should
pick the two suggestions, $s_1$ and $s_2$, so that we maximize 
the probability of any one of them being the intended program:
\begin{equation}
\argmax_{s_1,s_2}\, Pr(s_1 \cup s_2 \mid nl, D) \label{eqn:objective}
\end{equation}

When an LLM generates $N$ suggestions $L$ and we pick two program samples from the candidate list $L$ that have the highest average logprobs $\bar{p}$, it amounts to (approximately) trying to maximize: 
\begin{equation}
\argmax_{s_1,s_2}\, Pr(s_1 \mid nl, D) + Pr(s_2 \mid nl, D) \label{eqn:actual}
\end{equation}

This usually does not provide the user a good experience because this objective is not the same as Objective~\ref{eqn:objective}.
In fact, we know that 


\begin{eqnarray*}
\lefteqn{Pr(s_1 \cup s_2 \mid nl, D)} \\ & = & 
Pr(s_1 \mid nl, D) + 
Pr(\neg s_1\wedge s_2\mid nl, D) \\
& = & 
Pr(s_1 \mid nl, D) + 
Pr(s_2\mid nl, D) \cdot Pr(\neg s_1\mid s_2, nl, D) 
\end{eqnarray*}

The difference between Expression~\ref{eqn:actual} and the above expression is the factor $Pr(\neg s_1\mid s_2,nl,D)$. The probability $Pr(s_2\mid nl, D)$
is discounted by this extra factor in our desired objective. The probability $Pr(\neg s_1\mid s_2,nl,D)$ denotes the probability that the user does not accept $s_1$ given that they uttered $nl$, given $D$, and given that they would accept $s_2$. Now, if we assume that for any two programs $s_1,s_2$, the value $Pr(\neg s_1\mid s_2,nl,D)$ is close to $1$, then picking top-$K$ candidates by logprobs (Formula~\ref{eqn:actual}) would be good enough as it would be almost equal to our desired objective (Formula~\ref{eqn:objective}).
However, that assumption is unreasonable. In particular, if $s_1$ and $s_2$ both produce the same output on the user's input table $D$, then if the user accepts $s_2$, then they would very likely accept $s_1$ and hence $Pr(\neg s_1\mid s_2, nl)$ will be very small.

Let $[s]$ denote the equivalence class of program samples $s$ that consists of all
programs that produce the same output on user's input table $D$:
\begin{equation}
 [s] = \{ r \in PS \mid \text{exec } s(D) =  \text{exec } r(D) \}
\end{equation}
Our intuition is that if $[s_1] = [s_2]$, then $Pr(\neg s_1\mid s_2, nl, D)$ will be very small:
\begin{equation}
Pr(\neg s_1\mid s_2, nl, D) = \left\{
\begin{array}{cl} 1-\epsilon & \mbox{ if $[s_1] \neq [s_2]$} \\ \epsilon & \mbox{ if $[s_1] = [s_2]$}\end{array}\right\}\label{eqn:assumption}
\end{equation}
where $\epsilon$ is some small enough constant. 

If Assumption~\ref{eqn:assumption} holds, then Objective~\ref{eqn:objective} can be met by identifying the program $s_2$ such that $s_2$ is not in $[s_1]$ and $Pr(s_2 \mid nl)$ is maximal.
Semantic interleaving does just that:
we pick as $s_1$ the program with the largest probability, and then, as $s_2$, we pick the program with the largest probability that is {\em{not in the same equivalence class as $s_1$}}. We continue this process to pick the third and fourth sample as the programs with largest probability that are not in the same equivalence class as their predecessors. We do that until we run out of equivalence classes, and then go back and pick the second best program from each class. 

\begin{theorem}\label{thm:interleaving}
Let $Pr(s\mid nl, D)$ denote the probability of a user accepting program $s$ given the description $nl$ and input dataset $D$.  Assume that $\mathtt{score}$ is a function that assigns a number proportional to $Pr(s\mid nl, D)$ to every candidate $s$. 
Let $L\subset PS$ be a finite sample of possible candidate programs and $O$ be the outputs obtained by running
candidates in $L$ on the input dataset $D$.
If Assumption~\ref{eqn:assumption} holds, then the list $L^* := \mathsf{SemInterleave}({L, O, \mathtt{score}})$
of $K$ programs returned by $\mathsf{SemInterleave}$ satisfies the requirement
$$L^* = \argmax_{L'\subset L, |L'|=K} Pr(L \mid nl, D), \quad \forall 1\leq K\leq |\mathit{Distinct}(O)|.$$
\end{theorem}

When $K$ is greater-than the number of equivalence classes, then picking any $K$ candidates will involve pick multiple candidates from some equivalence classes. Under our assumption, the contribution of these candidates (picked from the same equivalence class as an existing candidate) to the desired Objective~\ref{eqn:objective} will be negligible irrespective of what candidates are picked. This completes our formal justification for semantic interleaving.

Theorem~\ref{thm:interleaving} shows that the initial set $L$ of candidates plays a crucial role -- it is critical to start with a {\em{diverse}} set $L$ so that we have a large number of equivalence classes to pick top-$K$ candidates. Theorem~\ref{thm:interleaving} also formally shows why the top-$K$ (ranked) candidates generated by LLMs are not the best candidates for showing to users, and why reranking is important.

\section{Advanced Techniques}
\label{sec:custom}

Semantic reranking, which includes semantic filtering and interleaving, is a powerful technique to improve the top-$K$ accuracy by leveraging the data context. In this section, we present two additional techniques that further enhance the improvements we get from semantic reranking. The advanced techniques are particularly effective for target languages that are not mainstream and that are not well represented in the training data for the LLM.


\subsection{Alternative Task Technique}

If an LLM struggles with a task -- such as, the task of generating M code from NL descriptions and input dataset -- then we could use a slightly different task to help us complete the original task.
The {\em{alternative task technique}} involves designing a task that is similar, but not identical, to the original task and then using the information from this alternative task to accomplish the original task.

Suppose we have a task to predict the random variable $Y$ given the random variable $X$; that is, given $X=x$ we want the model to produce the probability distribution $Pr(Y \mid X = x)$.
Suppose we can find an alternative random variable $Z$ (related to $Y$) such that we can estimate $Pr(Y=y,Z=z \mid X=x)$ extremely well. For example, $Y$ and $Z$ may be related to some deterministic (possibly reversible) function. In such a case, 
we can ask LLM to also predict $Pr(Z \mid X = x)$. Thereafter, we can take the samples $z$ and $y$ from $Pr(Z \mid X=x)$ and $Pr(Y \mid X=x)$ generated by the LLM respectively, and then use our knowledge of $Pr(Y=y,Z=z\mid X=x)$ to improve our prediction for $Y$.

The reason why this is better is because the LLM may use a "different sequence of neuron firings"
when it is tasked to predict $Z$, and thus more likely to give us new knowledge. 

\subsection{Output-Prediction based Score Tuning}
\label{sec:output_prediction}

Let us now instantiate the alternative task technique for our NL to code generation task. Here $X$ is defined by the NL description $nl$ and the dataset $D$. The random variable $Y$ is over the space of programs in the desired target language.
Our second random variable $Z$ will be over the output space of programs; that is, we want to predict the output that would be generated by a program when run on $D$.
We observed that the LLM \gpts is able to generate outputs with high fidelity from the NL utterance when the data context is available. In some cases, \gpts is able to predict the output directly when it is given the following prompt when we are targeting code in Power Query's M language:
\begin{quote}{\em{The assistant answers questions from a table by showing how the data is transformed in Power Query when given the description of the transformation task.}}\end{quote} 

\begin{algorithm}[t]
  \small
  \begin{algorithmic}[1]
    \Require An NL description $nl$, an input sample dataset $d$, candidates $L$, and their $\mathtt{logprobs}$, execution results $O_{\mathtt{exe}}$ for $L$, a function $\mathtt{sim}$ that measures similarity between two outputs s.t. it returns 1 when outputs are equal and value between $0$ and $1$ otherwise
    \Ensure Scores $\mathtt{scores}$ for each candidate in $L$
    \Function{OutputPrediction}{$nl$, $d$}
      \State $\mathsf{prompt} \gets \Call{PreparePromptForOutputPrediction}{nl, D}$
      \State $O_{\mathtt{pred}}, \mathtt{Ologprobs} \gets \Call{LLM}{\mathsf{prompt}, \mathtt{tmp}=0.6, N=25}$\label{line:predict-outputs}
      \State \Return $O_{\mathtt{pred}}, \mathtt{Ologprobs}$
    \EndFunction
    \Function{EstPr}{$s$, $o$, $O_{\mathtt{exe}}$}
      \State $o_{\mathtt{exe}} \gets O_{\mathtt{exe}}[s]$ \Comment{get output generated by $s$}
      \State $m \gets \mathtt{sim}(o, o_{\mathtt{exe}})$\label{line:estimate-pr}
      \State \Return $m$ \Comment{Estimate for the probability $Pr(s\mid o, nl, d)$}
    \EndFunction
    \Function{GetScores}{$L$, $\mathtt{logprobs}$, $O_{\mathtt{pred}}$, $\mathtt{Ologprobs}$, $O_{\mathtt{exe}}$}
      \For{$s \in L$}\Comment{for each candidate $s$}
        \State $\mathsf{score}[s] \gets \mathtt{logprobs}[s]$ \Comment{initialize score to avg. logprob}\label{line:score-start}
        \For{$o \in O_{\mathtt{pred}}$}\Comment{for each predicted output}
          \State $\mathsf{score}[s] \gets \mathsf{score}[s] + \mathtt{Ologprobs}[o] * \Call{EstPr}{s, o, O_{\mathtt{exe}}}$\label{line:score-end}
        \EndFor
      \EndFor
      \State \Return $O_{\mathtt{pred}}, \mathtt{Ologprobs}$
    \EndFunction
  \end{algorithmic}
  \caption{Output-prediction based Score Tuning}
  \label{algo:output-prediction}
\end{algorithm}

Recall that the semantic interleaving uses a score to order equivalence classes, and also order candidates within an equivalence class, which eventually plays a role in reranking and picking the top-$K$ candidates. The default score is simply the average logprob value provided by the LLM. We use the predicted output to further refine this score.
Our output-prediction based score tuning is shown in Algorithm~\ref{algo:output-prediction} and works as follows:
\\
(1) We use the LLM to predict 25 possible outputs $O_\mathtt{pred}$, along with their logprobs $\mathtt{Ologprobs}$, given $nl$ and $D$ (with temperature 0.6) on Line~\ref{line:predict-outputs}.
\\
(2) On Lines~\ref{line:score-start}-~\ref{line:score-end} we assign a new score to each candidate $s$ in the list $L$ of candidates as follows:
\begin{equation}
\mathtt{score}(p) = \mathtt{logprobs}(s) + \sum_{o\in O_{\mathtt{pred}}} \mathtt{Ologprobs}(o) * Pr(s \mid o, nl, D),\nonumber
\end{equation}
where $Pr(s\mid o, nl, D)$ is an estimate of the probability that $s$ is the desired program given $o$ is the desired output on $D$. \\
(3) We estimate this probability by using any similarity metric on the output space (on Line~\ref{line:estimate-pr}) to compare $o$ with the execution output $o_{\mathtt{exe}}$. The simplest metric is one that returns 1 if $o == o_{\mathtt{exe}}$ and 0 otherwise; however, one could use other metrics.

To relate back to alternative task technique, note that $\mathtt{Ologprobs}(o) \cdot Pr(s \mid o, nl, D)$ is an estimate of $Pr(s, o\mid nl, D)$, which is just the generic $Pr(Z=z, Y=y\mid X=x)$ used in the description of alternative task technique specialized to our case where $x$ is $(nl, D)$, $y$ is $s$ and $z$ is $o$.

Output prediction using LLMs can be accomplished in other ways too. Our overall methodology for using predicted outputs to rerank candidates is independent of {\em{how}} outputs are predicted. One could use the inputs $nl$ and $D$ to generate code in a different target (than what the user wants), and then execute that to predict outputs. That would be another instance of the alternative task technique, whose further investigation we leave for future work.

\subsection{Well-formedness based Score Tuning}
We add one further signal to the score assigned to a program $s$, namely the well-formedness of the output $o_{\mathtt{exe}}$ that would be generated by $s$. Recall that semantic filtering removes $s$ that fail to produce an output. However, even when a program $s$ succeeds to produce an output, certain outputs are less-likely to be the desired outputs. For example, if the output table contains a new column of null values, then we can mark this output as being less likely. Specifically, we scale the logprob value $\mathtt{logprobs}(s)$ assigned to a candidate $s$ by the model by a factor $\mathtt{DataQualityMetric}(o_\mathtt{exe})$. Again, we can use any data quality metric here, but in our experiments we use a simple one that penalizes $s$ if $o_{\mathtt{exe}}$ has null columns or if $o_{\mathtt{exe}}$ is an empty table. We also include well-formedness based score tuning inside ``output-prediction based score tuning'' for purposes of reporting experimental results. 




\subsection{Temperature Mixing}
\label{sec:boosting}


The techniques we previously introduced assume that the correct program is in the list of candidate programs generated by the LLM. However, in cases
where the LLM does not have sufficient prior knowledge of the data manipulation language, it is possible that the correct program is not included in the candidate set produced by the LLM. In this case, it makes sense to augment the candidate set with additional programs. This augmentation is done by leveraging the LLM itself and the fact that it performs temperature sampling~\cite{brown2020language}.

LLMs predict the next token by sampling from a  random variable. The sampling temperature is a hyperparameter that controls the randomness of the sampling process. The higher the sampling temperature, the more random the sampling process is. Concretely, if $V$ is the size of the vocabulary, then the $i$-th token is sampled with probability $\frac{e^{x_i/T}}{\sum_{j=1}^V e^{x_j/T}}$, where $x_j$ is the weight learned by the model for token $x_j$ and $T$ is the temperature. Temperature increases entropy.
If the temperature is high, the model can output, with rather high probability, tokens other than those with the highest $x_j$ values, making the generated output more diverse.

In the context of code generation, we noticed that the quality of programs synthesized from the LLM varies significantly with changing temperatures. In fact, there is a tradeoff. At higher temperatures, we get diverse $N$ samples, but the top-1 accuracy drops, and 
because the $N$ samples can exclude the one that has the highest average logprobs (e.g., the program that would be surfaced when temperature is set to 0).
On the other hand, at lower temperatures, we get the highest average logprobs candidate, but we lose diversity and the $N$ samples tend to contain the same candidate multiple times, resulting in top-$K$ accuracy for $K>1$ being very similar to the top-1 accuracy, which makes reranking unproductive. 

\begin{figure}[tb]
    \centering
    \includegraphics[width=0.9\linewidth]{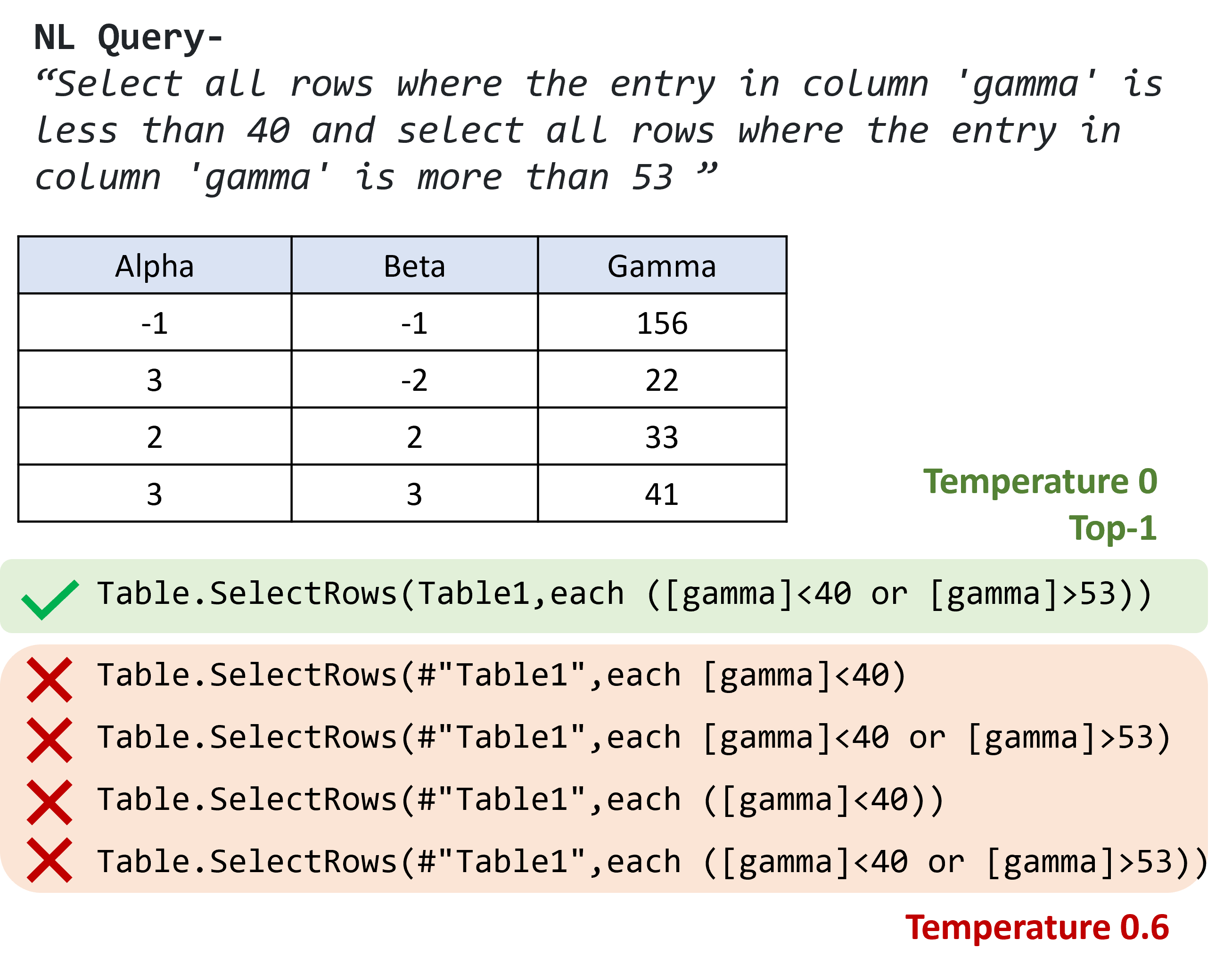}
    \caption{Example of \textit{temperature mixing}: The correct program is ranked second using logprobs at \textit{temperature 0.6}. Adding \textit{temperature 0} candidate bumps the correct program to the first rank.}
    \label{fig:boosting}
    \vspace{-2ex}
\end{figure}

To mitigate these issues and avoid missing correct programs, we introduce temperature mixing into our approach. In particular, we generate programs at both a low and a high temperature (i.e., 0 and 0.6, respectively), concatenate the results, as shown on Line~\ref{line:temp-mixing} of Algorithm~\ref{algo:overall}, and then apply the our reranking methodologies based on semantic filtering, interleaving, and output prediction. Temperature mixing is particularly effective when the model is more uncertain about the output, which can happen either because the query is ambiguous or very complex, or if the target language is unfamiliar to the model. In these cases, sampling at low temperature is important because the probability distribution computed by the model already has high entropy (more uncertainty) and lowering the temperature helps bring down the uncertainty. 
Candidates sampled at low temperatures are ranked higher, and hence temperature mixing can also help even when they add no new candidates; see Figure~\ref{fig:boosting}. In Section~\ref{sec:eval}, we discuss the performance improvements we get through this candidate augmentation process.



 





\section{Experimental Evaluation}
\label{sec:eval}
In this section we evaluate our approach in three data manipulation domains and provide an analysis of the experimental results. 

\subsection{Experimental Setup}

\subsubsection{The Targets}
We perform our evaluation on three different target languages:
SQL, the Power Query formula language M, and Python Pandas.
Our evaluation is performed on the task of generating code (in one of
these three targets) from natural language descriptions. Additionally, we assume
we are given an input dataset, and the goal
is to generate a formula or expression in the target language for the NL query 
that will work on the given dataset.

Both SQL and Pandas are popular languages and the pre-trained LLMs are expected
to have seen plenty of expressions in those languages in their training data, whereas M is a less popular language and LLMs may not have been exposed to as many M expressions in their training set.

\subsubsection{The LLM}
We use \texttt{code-davinci-002}, popularly known as the Codex model, for our evaluation. The model has since been replaced by newer and models, such as \texttt{text-davinci-003} and some new chat-based models; however, due to time, cost, and availability constraints, we performed our exhaustive evaluation on \texttt{code-davinci-002}. We executed some preliminary experiments on newer models and verified that the trends reported here also held true for the newer models.

As outlined in Section~\ref{subsec:setup}, LLMs expose hyper-parameters that influence their behavior. Our guiding principle in the design of our experiments was to pick the best possible choices for the hyper-parameters to define baselines (separately for each domain) and see if our proposed approach improves over it. For temperature, we used the setting $T=0.6$ or $T=0.8$ because these settings gave the best results 
in our baseline evaluations.

We ask the model to generate $N=25$ candidates typically, and then unless otherwise stated, the baseline picks the top-$K$, for whatever value of $K$ we are considering, based on the average logprobs of the $25$ solutions. \eat{Some of the newer LLMs have ceased to provide the logprobs of the generated tokens, in which case, the default order returned by the model is used to select the top-$K$ candidates.}
The stop sequence for the LLM is set to the token that is identified in the prompt as marking the
end of a SQL, M, or Pandas expression.
All the other hyperparameters of the model take their default values as mentioned in~\cite{api}.

\subsubsection{The Benchmarks}

For evaluating our approach for Pandas, we used  
the ``Jigsaw'' dataset~\cite{jain2022jigsaw}. 
Since the M expression language is limited and there are not any available public benchmarks, we leveraged the Jigsaw dataset to create a benchmark for M.
In particular, we filtered the Jigsaw dataset and extracted only
the transformations that M supports to create the ``\JigsawM'' benchmark set for M. 
Additionally, we also used two other datasets: The first one was created by us by scraping
PowerQuery help forums and collecting M expressions' NL descriptions
and M expressions from there, which we call ``\Forum'' in the tables. The second one was a set of 
benchmarks we obtained from the PowerQuery team and is called 
``\Product'' in the tables.
For SQL, we used the ``Spider'' dev~\cite{yu2018spider} and
``KaggleDBQA''~\cite{lee2021kaggledbqa} datasets. 


\begin{table}[t]
\caption{Dataset Statistics}\label{table:data-statistics}
\begin{tabular}{lllll} \toprule
Target & Benchmark &  \# NL  & Avg NL& Avg Code \\
 & & Questions & Question & Length  \\
 & &  & Length &   \\
\midrule
SQL & Spider & 1034 & 68.04 & 108.32 \\
SQL & KaggleDBQA & 272 & 55.79 & 96.00 \\
M &\Forum & 25 & 88.28 & 96.96 \\ 
M &\Product & 34 & 38.08 & 68.94 \\ 
M & \JigsawM & 442 & 65.66 & 75.83 \\
Pandas & Jigsaw & 793 & 70.49 & 56.47 \\
\bottomrule
\end{tabular}
\end{table}

Table~\ref{table:data-statistics} presents some statistics about the benchmarks. Each benchmark consists of a set of NL statements and associated code pairs. As shown in the table, we evaluated our techniques on a total of $2526$ (NL, code) pairs. In particular, we used a total of 1306 queries for SQL, 501 queries for M and 793 queries for Pandas. The average number of characters in the NL description and code are given in the last two columns of Table~\ref{table:data-statistics}.

\subsubsection{Metrics} We use {\em{execution match accuracy}} as the metric for evaluation. A candidate execution matches the ground truth if both programs return identical outputs when run on the input dataset. We report the percent of benchmarks where we get an execution match in our tables below. We also collected {\em{exact match accuracy}} -- where we test if the candidate syntactically matches the ground truth. Those numbers are smaller, but the improvement trends are similar for either metric and hence we focus on one in the presentation below.

\subsection{Experimental Results}
We first start by presenting the net improvements we get over the baseline by using all our
techniques combined (semantic reranking and output-based approaches) as applicable
for each target and benchmark class. Table~\ref{table:overall} shows these results.
We report the execution match accuracy (percentage) we get at $K \in \{1,3,5 \}$ for the different benchmark sets. The baseline is obtained by turning off the additional steps added by our approach. The {\em{absolute}} improvement we see over the baseline numbers is reported within parenthesis in Table~\ref{table:overall}. By {\em{absolute}} we mean that the baseline accuracy is the number outside the parenthesis minus the number within.

Note that the baseline approach uses the same prompt as our approach. The prompt includes five few-shot examples, input table name and schema,  and at least three sample rows (three). In other words, the baseline uses the \textit{best possible prompt} we could design for the task. 
We did not consider as baseline any approaches that utilize custom ML models or require fine-tuning large language models (see Section ~\ref{sec:related}) as we don't want to make any assumption about availability of training data. It is also worth noting that our baseline already surpasses the SOTA using prompt engineering~\cite{pourreza2023dinsql} as depicted in the Spider leaderboard~\cite{spiderleaderboard}.

\begin{table}[t]
    \caption{The Top-1,3,5 execution match accuracy obtained using our approach along with the gains (in brackets) over corresponding baselines.}
    \begin{tabular}{llccc} \toprule
    \multicolumn{2}{c}{Task} & \multicolumn{3}{c}{Execution Match Accuracy @K} \\
    \cmidrule(lr){1-2} \cmidrule(lr){3-5}
     {Target} & {Benchmark}  & {K=1} & {K=3} & {K=5} \\ \midrule
SQL & Spider & 76.0 (+02.8) & 90.5 (+12.1) & 92.8 (+11.9) \\
SQL & KaggleDBQA & 63.2 (+00.5) & 78.4 (+12.5) & 81.1 (+12.5) \\
M & \Forum & 61.6 (+17.6) & 74.4 (+19.2) & 74.4 (+10.4) \\
M & \Product & 72.3 (+10.6) & 75.3 (+02.4) & 76.5 (+00.6) \\
M & \JigsawM & 64.7 (+45.0) & 72.52 (+33.8) & 73.7 (+25.7) \\
Pandas & Jigsaw & 74.1 (+02.9) & 86.8 (+13.4) & 89.0 (+12.9) \\
        \bottomrule
    \end{tabular}    
    \label{table:overall}
\end{table}


For $K=1$, we see improvements in the range $0.5\%$ (for SQL on the Kaggle dataset) to $45\%$ (for M on \JigsawM dataset). Improvement at Top-1 are significantly higher for M than for SQL and Pandas because without our enhancements, the LLM struggles on an unfamiliar language like M.
We see the benefits of our approach on ranking when $K=3$ and $K=5$: our techniques bring the desired solution closer to the top.
For $K=3$, we see improvements consistently more than $12\%$ (with the exception for M on \Product).
For $K=5$, we see improvements consistently more than $10\%$ (with the exception for M on \Product again).
The gains on \Product benchmarks are smaller since those benchmarks are smaller and simpler: 
Table~\ref{table:data-statistics} shows that the average length of both the NL question and the ground-truth code is smallest for \Product; and baseline accuracy at $K=1$ is highest for \Product (at $61.7$) among all M tasks.

We will now present the gains
from each of the components of our overall approach.
We start with the baseline, then enable each component
one by one. We report the accuracy achieved in each step (and hence,
the improvement over the previous step).

\medskip
\noindent
{\bf{Improvement from Semantic Filtering}}

\smallskip
\noindent
To measure the improvement from semantic filtering, we first ask the model to generate $25$ responses for the NL question at hand. We then take the ordered list of candidates generated by the model and execute them on a sample of the input tables. If the execution is successful, we say that the candidate is \emph{semantically valid}. We only keep the candidates that are semantically valid (while preserving the initial ordering) and report the accuracy on the first $K$ candidates in Table~\ref{table:filtering}. 

Table~\ref{table:filtering} shows the gains we get by adding semantic filtering to the output of LLM. 
We compare the execution match accuracy at $K$, for 
$K \in \{1,3,5\}$ observed when we add semantic filtering to the
baseline. The baseline does not use semantic filtering {\em{but keeps 
everything else the same}} (i.e, prompts, hyperparameters, etc.).

\begin{table}[t]
    \caption{Gains from Adding Semantic Filtering.}
    \begin{tabular}{llccc} \toprule
    \multicolumn{2}{c}{Task} & \multicolumn{3}{c}{Execution Match Accuracy @K} \\
    \cmidrule(lr){1-2} \cmidrule(lr){3-5}
     {Target} & {Benchmark}  & {K=1} & {K=3} & {K=5} \\ \midrule
SQL & Spider & 76.0 (+2.8) & 80.5 (+2.1) & 82.8 (+1.9) \\
SQL & KaggleDBQA & 63.2 (+0.5) & 66.5 (+0.6) & 68.6 (+0.0) \\
M & \Forum & 51.2 (+7.2) & 64.8 (+9.6) & 69.6 (+5.6) \\
M & \Product & 71.2 (+9.5) & 75.3 (+2.4) & 76.5 (+0.6) \\
M & Jigsaw & 39.5 (+19.8) & 55.1 (+16.4) & 61.9 (+14.0) \\
Pandas & Jigsaw & 74.1 (+2.9) & 75.7 (+2.3) & 78.1 (+2.0) \\
        \bottomrule
    \end{tabular}    
    
    \label{table:filtering}
\end{table}

    

There is never a drop in execution match accuracy because
semantic filtering only removes candidates that would definitely
fail the execution match check. While we observe decent gains
across the board, the gains are more profound for M. This
is because the model is less familiar with the M language
and hence is more likely to generate M expressions that would
not successfully execute.

The average number of candidate programs remaining after the semantic filtering step is $8.4$ for the Spider dataset, $6.3$ for the KaggleDBQA dataset,  $8.3$ for the three M datasets, and $5$ for the Jigsaw Pandas dataset. That is a significant reduction from $25$ in every case. The maximum number of programs filtered out is 24/25 and the minimum is 0. The errors encountered during program execution include erroneous column and table names (i.e., column \texttt{operationFrom} is written as \texttt{operation\_from}) and extra or missing characters in the queries.
Note that semantic filtering removes both syntactically incorrect programs, as well as programs that parse, but throw run-time errors. 



\medskip
\noindent
{\bf{Improvement from Semantic Interleaving}}

\smallskip
\noindent
To measure the improvement from interleaving, we take the semantically valid 
candidates generated above and provide that as input to our interleaving-based re-ranking function,producing a different ordering of the candidates. 
We report the accuracy on the first $K$ candidates in Table~\ref{table:interleaving}.
%
Note that the improvement numbers
(within brackets) reported in Table~\ref{table:interleaving} are 
gains {\em{in absolute terms}} over the numbers reported in 
Table~\ref{table:filtering}.
We note that semantic interleaving does not change the top candidate and hence it does not influence the
Top-1 accuracy, and hence we only show
execution match accuracy at $K=3$ and $K=5$.

\begin{table}[t]
    \caption{Gains from Adding Semantic Interleaving.}
    \begin{tabular}{llcc} \toprule
    \multicolumn{2}{c}{Task} & \multicolumn{2}{c}{Execution Match Accuracy @K} \\
    \cmidrule(lr){1-2} \cmidrule(lr){3-4}
     {Target} & {Benchmark}  & {K=3} & {K=5} \\ \midrule
SQL & Spider & 90.5 (+10.1) & 92.84 (+10.1)
\\
SQL & KaggleDBQA & 78.4 (+11.9) & 81.1 (+12.5)
\\
M & \Forum & 67.2 (+2.4) & 70.4 (+0.8)
\\
M & \Product & 75.3 (+0.0) & 76.5 (+0.0)
\\
M & Jigsaw & 60.5 (+5.4) & 67.9 (+6.0)
\\
Pandas & Jigsaw & 86.8 (+11.1) & 89.0 (+10.9)
\\
        \bottomrule
    \end{tabular}    
    
    \label{table:interleaving}
\end{table}

    

We observe that semantic interleaving provides around 10\%
absolute gain in semantic match accuracy for Pandas and SQL,
whereas the gain is generally smaller for M. 
This is almost the reverse of what we observed for
semantic filtering where the gains were higher for $M$ than for
Pandas and SQL.
This can be interpreted as follows: for targets Pandas
and SQL that
are reasonably well represented on the web,
LLMs do not have difficulty with generating syntactically
correct and executable expressions, but have some 
difficulty with ranking them correctly; and the situation
is reversed for targets like M that are not as well represented
in the web data.


The gains reported in Table~\ref{table:filtering} and Table~\ref{table:interleaving}
are additive, so the total gain from using semantic filtering and interleaving is more than
10\% in absolute terms for most benchmark classes and goes as high as 21.8\% for M on \JigsawM 
for $K=3$.

\medskip
\noindent
{\bf{Improvement from Output-based Score Tuning}}

\smallskip
\noindent
We next evaluate the improvements we get using output-based score tuning, which includes both output-prediction based tuning and well-formedness based tuning.
As discussed earlier, this technique was designed for target 
languages that are not well-represented in the LLM's training data.
Hence, we only report the numbers here for M, and note that the we did not
observe any significant gain using this technique on Pandas and SQL.

\begin{table}[t]
    \caption{Gains from Output-based Score Tuning (M only).}
    \begin{tabular}{llccc} \toprule
    \multicolumn{2}{c}{Task} & \multicolumn{3}{c}{Execution Match Accuracy @K} \\
    \cmidrule(lr){1-2} \cmidrule(lr){3-5}
     {Target} & {Benchmark}  & {K=1} & {K=3} & {K=5} \\ \midrule
M & \Forum & 60.0 (+8.8) & 73.6 (+6.4) & 74.4 (+4.0)
\\
M & \Product & 71.2 (+0.0) & 75.3 (+0.0) & 76.5 (+0.0)
\\
M & \JigsawM & 61.3 (+21.8) & 72.5 (+12.0) & 73.7 (+5.8)
\\
        \bottomrule
    \end{tabular}    
    
    \label{table:output-ranking}
\end{table}

    

Table~\ref{table:output-ranking} reports execution accuracy observed on
M benchmarks for $K \in \{1,3,5\}$ when we additionally add output-based ranking.
The numbers in parenthesis report the improvement over the accuracy
numbers reported in Table~\ref{table:interleaving} where we did not use
output-based ranking.
Output-based ranking is able to significantly
 improve Top-1 accuracy for 2 out of the 3 benchmark classes for M;
in fact, up to $21.8\%$.
This matches the intuition that output-based ranking can potentially
exploit alternate new pathways of the LLM  to help generate potential candidates,
which is especially helpful for benchmarks where the direct use of
LLM yields poor results. 
%
It is able to bring Top-1 accuracy to the 60\%-70\% for all classes.
We see no improvement in the \Product class where the Top-1 accuracy, most likely due to reasons mentioned earlier. 
We see output-based score tuning benefit 2 and 95 benchmarks at K=1 from \Forum and \JigsawM respectively.

We observe gains for $K \in \{3,5\}$, although those gains are more modest
compared to gains at $K=1$. Nevertheless, even for $K \in \{3,5\}$, 
output-based score tuning tends to do whatever is necessary to get the accuracy to around $75\%$ range.

\medskip
\noindent
{\bf{Improvement from Temperature Mixing}}

\smallskip
\noindent
We next evaluate the gains from temperature mixing. This is also a technique that helps for languages such as $M$ that are not well-represented in LLM's training data.
Temperature mixing only adds one candidate from the temperature-0
run, and hence it typically only influences the Top-1 accuracy.
Hence, in Table~\ref{table:temp-boosting} we only report numbers for
$K=1$ for M.  The accuracy for $K=3,5$ was identical to
the accuracy in Table~\ref{table:output-ranking} and hence we
get $0\%$ gain for those cases. 
However, for $K=1$, we see some gains in the range
$0\%$ to $3.8\%$ in
execution match accuracy. 
While the gains may seem insignificant,
they are very useful since they improve Top-1 accuracy.

\begin{table}[t]
\caption{Gains from Temperature Mixing.}
    \begin{tabular}{llc} \toprule
    \multicolumn{2}{c}{Task} & \multicolumn{1}{c}{Execution Match Accuracy @K} \\
    \cmidrule(lr){1-2} \cmidrule(lr){3-3}
     {Target} & {Benchmark}  & {K=1}  \\ \midrule
M & \Forum & 61.6 (+1.6) 
\\
M & \Product & 72.3 (+1.1) 
\\
M & Jigsaw & 64.7 (+3.4) 
\\
        \bottomrule
    \end{tabular}    
    
    \label{table:temp-boosting}
\end{table}

\section{Discussion and Future Work}


A key assumption underlying several of our reranking techniques is that candidates generated by the LLM can be executed inside a try-catch block.
%
This assumption is easy to satisfy for languages that have few or no side-effects. This is the case for the PowerQuery M target language. For such languages, we can use execution-based interleaving in production. 
However, when the language is richer and more general purpose, such as Python, models like \gpts can recommend programs that have negative side-effects (e.g., deleting the operating system, etc). In this case, we have two options. The first option is to execute untrusted code from the LLM in a sandbox. A second, cheaper, alternative is to design non-execution based techniques that try to approximate the execution-based techniques.

\begin{table}[t]
\caption{Gains from semantic interleaving with and without execution on the Jigsaw benchmark.}
    \begin{tabular}{lccc} \toprule
    \multicolumn{2}{c}{Task} & \multicolumn{2}{c}{Execution Match Accuracy @K} \\
    \cmidrule(lr){1-2} \cmidrule(lr){3-4}
     {Target} & {Execution-Based}  & {K=3} & {K=5} \\ \midrule
Pandas & Yes & 86.8 (+11.1) & 89.0 (+10.9) \\
Pandas & No  & 79.1 (+3.4) &  81.4 (+3.3) \\
        \bottomrule
    \end{tabular}    
    
    \label{table:pandas_exec_results}
\end{table}

We have done some initial investigation towards this direction in the context of NL to Pandas. The idea is to compute equivalence classes by clustering the candidates (rather than by executing them). So, we generate features based on the program syntax that we use to perform program clustering. We developed a custom ANTLR4 \cite{parr2013definitive} parser that provides a logical representation for any Python query that leverages the Pandas library. We use the features laid out by the custom parser (e.g., Pandas operators) to group code snippets into their respective classes. Our results, shown in Table~\ref{table:pandas_exec_results}, indicate that we can retrieve about $3.3\%$ gains by clustering, out of the full $11\%$ that we achieved with execution. There is clearly room for improvement here.
Specifically, the intriguing challenge in the domain of code generation and interleaving lies in the ability to quickly establish the semantic equivalence of two \eat{(SQL, M, or Pandas)} queries as most applications of NL to code have low-latency requirements.

\section{related work}
\label{sec:related}


\subsection{Few-shot Prompting}
Our contributions are not related to few-shot prompting, but we exploit them for building our baseline.
Few-shot prompting refers to inclusion of some concrete examples of the task in the prompt. It has been shown to help the LLM generate good program recommendations~\cite{brown2020language, chen2022program, chen2021evaluating, nijkamp2022codegen}, 
including recommendations in less popular languages~\cite{hendy2023good}. A wide collection of work exists on few-shot prompting ranging from crafting prompt templates \cite{shin2020autoprompt, zhong2021factual, gao2020making, shi2022natural}, considering the permutations of examples \cite{zhao2021calibrate, lu2021fantastically}, to increasing the number of few-shot examples \cite{wei2022chain}. We build our baselines using these references, and we select exemplars to include in our few-shot prompt by using the popularized KATE (Knn-Augmented in-conText Example selection) method \cite{liu2021makes}: an embedding model is used to convert the NL queries into a vector representation, 
and $k$ examples are retrieved from a knowledge base that are nearest neighbors of the user's NL query in the embedding space.
Given LLM's sensitivity to prompts, many works exist in prompt aggregation\cite{arora2022ask}, or training models that perform aggregations itself \cite{jiang2020can, schick2020s}, as well as chain-of-thought prompting \cite{liu2023comprehensive}, and, more recently, repair \cite{chen2023teaching, shinn2023reflexion}, but we leave these as potential directions for future work.

\subsection{Data Context}
Since we are operating in the domain in which \textit{data is available}, we tested various ways to summarize the associated input data in the prompt as it is well-known that small changes in the prompt can have significant effects on the generated programs~\cite{min2022rethinking}. Examples include using encoding the input data within \texttt{CREATE SQL} statements, introducing new tokens like \texttt{<T>} for demarking table names
\eat{, \texttt{<C>} for representing columns, and \texttt{<S>} for primary and foreign keys,} as well as, simple dictionaries that list each table and its associated column attributes and types \cite{scholak2021picard, shaw2020compositional}.  
Similar to existing work \cite{gemmell2023generate}, we include a sample of $3$-$8$ rows per table in the prompt. Our main contribution is about using data context for post-processing, but we do summarize data in the prompts to define the baseline.
\eat{For the KaggleDBQA and Jigsaw benchmarks, this threshold was often sufficient to capture the full input data table; but this is not the case for the Spider dataset where this cut-off resulted in many tuples not surfaced to the LLM. }

\subsection{Natural Language to Code} In the context of databases, early work on building natural language interfaces involved leveraging ontologies, intermediate languages and various heuristics for join path selection~\cite{athena, athena++, nalir}. With the advent of transformers~\cite{vaswani2017attention}, it has become much simpler to provide such capabilities on top of a relational database. The Spider leaderboard~\cite{spiderleaderboard} contains a list of works that leverage machine learning for text-to-SQL generation and are evaluated on the Spider dataset. The approaches fit into three categories: custom ML models (e.g.,~\cite{li2023resdsql, catsql, cao-etal-2021-lgesql, xu-etal-2021-optimizing}), prompt engineering with pre-trained language models such as Codex and GPT-4~\cite{pourreza2023dinsql, synchromesh}, and fine-tuned large language models~\cite{picard, shaw-etal-2021-compositional}. Our work falls into the second category as we operate under the assumption that we do not have enough data to train a custom model or to fine-tune a large language model. The top performance results in this category are obtained by the work in~\cite{pourreza2023dinsql}. This work achieves $74.2\%$ and $69.9\%$ top-1 execution accuracy on the Spider dev test (the dataset we are also using for our evaluations) using the GPT-4 and Codex models respectively. Our approach provides $76\%$ top-1 execution accuracy using the Codex model demonstrating that we are able to surpass the SOTA methods using prompt engineering and LLMs. Moreover, our new techniques contribute significantly to improving the top-$K$.

In the context of Pandas, the most relevant work to ours is the one published in~\cite{jain2022jigsaw}. The main contrast lies in the fact that their method necessitates input/output test cases from the user. These tests are used to validate and refine the programs generated by the LLM, or to modify the LLM-produced code so that it can satisfy the test cases. In contrast, our method solely relies on the natural language utterance and does not require any additional input. 

\subsection{Reranking}
Generating code from natural language is challenging \cite{yu2018spider, chen2021evaluating, austin2021program, li2022competition}. Since the desired code is more likely to be generated when multiple programs are sampled, there is extensive work around designing reranking techniques, including execution-based reranking techniques, to select the best candidate among multiple samples \cite{shi2022natural, zhang2022coder, ni2023lever, li2022competition}. However, most work has focused on improving Top-1 accuracy~\cite{shi2022natural,ni2023lever,zhang2022coder}, whereas we focus on techniques for top-$K$ improvements. Unlike our work, some works consider a different signal for reranking: namely, translating the code back the NL and checking consistency, which is related to maximizing mutual information objective to pick the top candidate~\cite{back-translation,li-etal-2016-diversity,zhang2022coder}, which we can integrate in our score-based reranking framework.
There is recent work~\cite{li2022competition} that considers techniques similar to ours to improve Top-K accuracy; however, it is heavily targeted to competition-level coding, and it works at a different scale that is unrealistic for actual applications; for e.g.,
it generates order of hundred thousand samples, filters them down to thousands, and then clusters
and picks 10 by interleaving. A crucial distinction with all previous work, including~\cite{li2022competition}, is that our contributions are not just empirical and we provide theoretical justification for interleaving.
Moreover, we introduce the new alternative task technique, and its instantiation to output-prediction based score tuning.


\subsection{LLMs for Data Management} In-context learning is a relatively new concept, and there is limited research on its potential benefits for data processing. Chen et al.~\cite{chen2023symphony} proposed a system for querying heterogeneous data lakes with in-context learning. An alternate approach requiring lesser space for more preprocessing was proposed by~\cite{arora2023language}. In-context learning has also been applied to data wrangling~\cite{narayan2022can} and processing SQL queries~\cite{trummer2022codexdb}, but these require manual prompt design, which can be a challenge for data management systems due to the variety of formats, attribute-types, and topics found in documents. Most approaches using LLMs are limited to single tables.~\cite{vogel2022towards} made the observation that there is significant potential to learn from the full structure of the relational database, including neighboring tables that can contain important information for a contextualized representation. Finally,~\cite{vos2022towards} conducted a preliminary study on how to scale data wrangling with LLMs for data integration and cleaning tasks. They observed that prompting is attractive for such use cases as it can re-use a single pre-trained model for several tasks and tables. However, it requires high expertise and manual effort to engineer high-quality task- and data-specific prompts, which is not feasible for enterprise databases with thousands of different tables. Finetuning on the other hand incurs low manual costs but high storage costs. To achieve the best of both worlds, they use \emph{prefix tuning}~\cite{li2021prefix} as a parameter-efficient alternative to finetuning for data-wrangling tasks. A similar approach may also be beneficial for other NL to X tasks as well.





\section{conclusion}
\label{sec:conclusion}

In this paper, we presented a novel program synthesis framework for data manipulation programs based on in-context learning. Our approach leverages the data context end-to-end by exploiting both task inputs and outputs in conjunction. We evaluate our framework in three different domains (databases, data science, business intelligence) using a variety of new and existing benchmarks. Our results highlight substantial improvements in top-$K$ accuracy across all three domains.

\bibliographystyle{ACM-Reference-Format}
\bibliography{ref}

\end{document}